\documentclass[final,11pt]{elsarticle}
\usepackage[numbers]{natbib}  

\usepackage{amssymb}
\usepackage{amsmath}
\usepackage{units}
\usepackage[printonlyused]{acronym}
\usepackage[hidelinks]{hyperref}
\usepackage[left=2.5cm, right=2.5cm, top=3.5cm, bottom=2.5cm]{geometry}

\newcommand{\Aeff}{A_{\text{eff}}}

\newcommand{\Latt}{L_{\text{att}}}
\newcommand{\amp}{\mathcal{A}}
\newcommand{\phase}{\Delta\varphi}

\newcommand{\df}{\Delta\nu}
\newcommand{\Abs}{\mathcal{A}}
\newcommand{\vgair}{v_{g,\text{air}}}

\newcommand{\picsize}{0.7\textwidth}

\journal{Nuclear Inst. and Methods in Physics Research, A}

\begin{document}

\begin{frontmatter}

\title{A novel method for measuring the attenuation length and the group velocity of transparent liquids in a variable length cavity}

\author{Jessica Eck} 
\author{Dhanushka Bandara}
\author{Gina Grünauer}
\author{Tobias Heinz}
\author{Tobias Lachenmaier}

\affiliation{organization={Physikalisches Institut, Eberhard Karls Universität Tübingen},
            addressline={Auf der Morgenstelle 14}, 
            city={Tübingen},
            postcode={72076}, 
            country={Germany}}

\begin{abstract}
The transparency of liquid scintillators or water is an important parameter for many detectors in particle and astroparticle physics. In this work, the \ac{CELLPALS} method for the determination of the attenuation length is presented for the first time. The method is based on an experimental setup similar to a Fabry-P\'erot interferometer but adding up multiple-reflected intensities of a modulated light source. \ac{CELLPALS} was developed to measure the attenuation length of highly transparent liquids ($>\unit[10]{m}$) with significantly lower uncertainties than with UV-Vis spectroscopy, which is a standard method for determining the attenuation length. In addition to the attenuation length, the group velocity of light in the sample can also be derived from the free spectral range of the cavity, which is not provided by any conventional method for determining the attenuation length. In this work, the \acs{CELLPALS} method, its achievable precision and an experimental setup to demonstrate its feasibility are discussed. The attenuation lengths and group velocities of several transparent liquids were measured at wavelengths between $\unit[420]{nm}$ and $\unit[435]{nm}$. In addition, the attenuation length and group velocity of \ac{LAB} samples after different purification stages and a purified \acs{LAB}-based \acl{LS} were measured at a wavelength of $\unit[425]{nm}$. The results confirmed the potential of \acs{CELLPALS} to determine the attenuation length with an uncertainty of $\sim\unit[2]{\%}$. The group velocity of light in the sample can be determined with an uncertainty of $\sim\unit[0.03]{\%}$. 
\end{abstract}


\end{frontmatter}

\section{Introduction}
\label{sec:Introduction}

Liquid scintillator (\acs{LS}) and water Cherenkov detectors are used in particle physics, in particular for neutrino experiments. Prominent representatives of such large volume neutrino detectors are
KamLand~\cite{KamLAND}, Borexino~\cite{Borexino}, SNO+~\cite{SNO} , \acs{JUNO}~\cite{JUNO}, Super-Kamiokande~\cite{SuperK} and Hyper-Kamiokande~\cite{HyperK}. Since the cross-section of neutrino interactions is usually very small, a large
amount of liquid is required. Photons generated by scintillation or the Cherenkov effect typically travel distances of several tens of metres to reach the photo sensors. In order to achieve sufficient transparency (e.g.~$>\unit[20]{m}$ for \acs{JUNO}~\cite{JUNO}), elaborate purification systems are installed~\cite{PurificationPlantJUNO}, and in this context, suitable methods for measuring large attenuation lengths are used for monitoring the
individual steps. The \acf{CELLPALS} method enables the measurement of such large attenuation lengths and also provides the group velocity of the sample, which is useful for improving the spatial reconstruction of events.

The transparency of a medium is described by its attenuation length $\Latt$, which, according to Beers' law
\begin{equation}
 I(L)=I_0\exp\left(-\frac{L}{\Latt}\right)\label{eq:Beer}
\end{equation}
denotes the distance $L$, which leads to a reduction of the light intensity to $1/e$ of the initial intensity $I_0$. The attenuation length $\Latt^{-1}=L_{A}^{-1}+L_{s}^{-1}$ describes the total effect of attenuation  taking
into account the absorption length $L_{A}$ and scattering length $L_{S}$~\cite{ScatteringLength}.

One method for determining the attenuation length of liquids is UV--Vis spectroscopy by measuring absorption spectra in the ultraviolet and visible light range~\cite{Absorbance,UV-VIS_Method}. Using a monochromator, this
method is suitable for wavelength-dependent monitoring of the absorbance. However, since the standard cuvettes containing the samples have a maximum length of $L=\unit[10]{cm}$, the attenuation by the liquid is less than
$\unit[0.5]{\%}$ for highly transparent media with $\Latt>\unit[20]{m}$. Considering the losses due to reflections $\sim\unit[7]{\%}$ that occur at the boundary surfaces between the air and the cuvette, the determination of large attenuation
lengths by the UV--Vis method is affected by large uncertainties (cf.  \ref{sec:UV_Vis}).

In order to determine high attenuation lengths more precisely, a number of other methods have been developed, which are mainly based on increasing the light path. These devices typically use a vertical tube with a length
of $\sim\unit[(1-3)]{m}$ as a sample holder with a variable fill level. By measuring the intensity as a function of fill height, it is possible to derive the attenuation
length~\cite{PALM,Latt_method1,Latt_method2,Latt_method3,Latt_method4}. 

The \ac{CELLPALS} method allows the determination of the attenuation length of transparent liquids with an uncertainty comparable to these methods and with a significantly lower uncertainty than the UV--Vis method. Due to its compact design, the \ac{CELLPALS} method requires only a small sample volume ($\sim\unit[1]{litre}$) and is very straightforward in terms of cleaning and changing the sample. The location of the \ac{CELLPALS} setup can also be easily changed, making it ideal as a direct monitoring system at the purification plant. Similar to the UV--Vis method, \ac{CELLPALS} also compares the intensities of a reference beam and a sample beam, but the fundamental difference of \ac{CELLPALS} is the extension of the effective light path by using an optical cavity in combination with an amplitude-modulated laser as the light source. The \ac{CELLPALS} method is based on Fabry-P\'erot interferometry, where it is not the wavelength of the light that interferes, but the sinusoidally modulated part of the intensity (cf. \ref{sec:Mechanism of CELLPALS technique}). In \ref{cap:CELLPALS} the frequency-dependent analysis of the attenuated intensity is introduced which is used to derive the attenuation length $\Latt$ and the group velocity $v_g$ of the sample. The achievable precision of the \ac{CELLPALS} method is discussed in \ref{cap:precision} and the experimental setup is shown in \ref{sec:Exp_setup}. In \ref{sec:UV_Vis}, the UV--Vis method for the determination of the attenuation length is briefly described and in \ref{sec:Measurements}, the measurements with \ac{CELLPALS} are presented together with a comparison to the measurements with the UV--Vis method. Finally, all the results obtained in this work are summarised and discussed in \ref{sec:Conclusion}.

\section{The CELLPALS principle}
\label{cap:CELLPALS}

\begin{figure}[htb]
	\centering
	\includegraphics[width=\picsize]{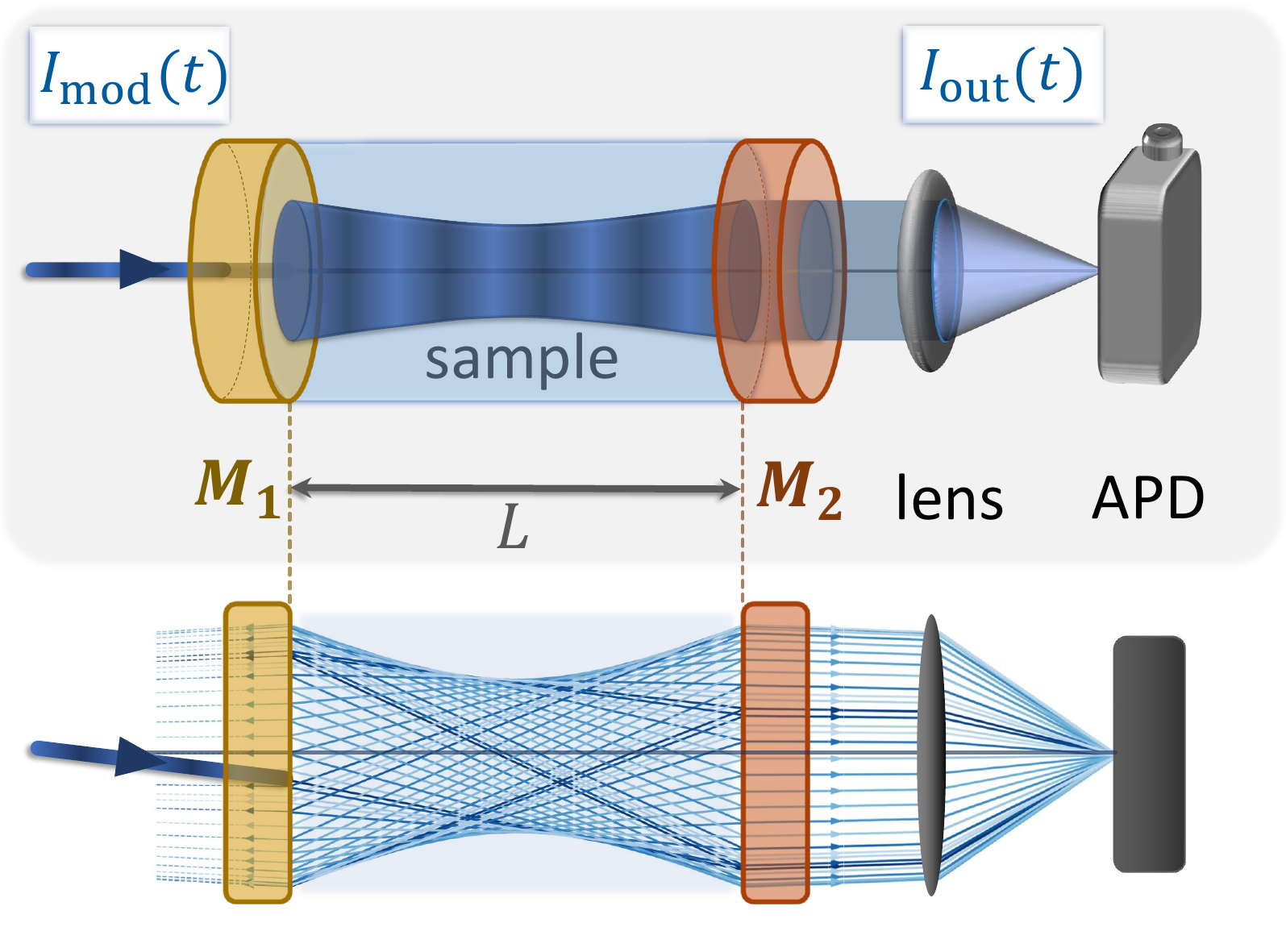}
	\caption{Schematic representation of the resonator of length $L$ defined by two semi-transparent concave mirrors $M_1$ and $M_2$ containing the sample. The initial modulated intensity $I_{\text{mod}}(t)$ is attenuated inside the cavity and the outgoing intensity $I_{\text{out}}(t)$ is collimated by a lens and detected by an \acs{APD}. In the lower plot, the beam inside the cavity is illustrated using Gaussian beam propagation and the outgoing intensity $I_{\text{out}}(t)$ is represented as a superposition of an infinite number of individual rays $I_n(t)$. The rays have been attenuated to different degrees (indicated by the brightness of the rays) and have different phase shifts in their modulation depending on their number of round trips $n$.}
	\label{fig:Resonator}
\end{figure}

The \ac{CELLPALS} principle is based on the attenuation of the intensity according to Beers' Law (cf. Eq. \ref{eq:Beer}) and uses the principle of Fabry-P\'erot interferometry. An optical resonator of length
$L$ is used to increase the effective light path $L_{\text{eff}}=\kappa\cdot L$ through the sample by the factor $\kappa=\left(1-R\right)^{-1}$~\cite{CEAS}.
This factor depends on the reflectivity $R$ of the two opposing semi-transparent concave mirrors $M_1$ and $M_2$, which define the cavity resonator containing the sample (cf.  Fig. \ref{fig:Resonator}).
After entering the resonator through $M_1$, the light is reflected back and forth between the two mirrors. The total intensity $I_{\text{out}}(t)$ leaving the cavity resonator behind mirror $M_2$ is detected as a signal.

\subsection{Detectable intensity after the cavity resonator}
\label{sec:Mechanism of CELLPALS technique}

The intensity emitted by the laser diode can be described by $I_{\text{laser}}(t)=I_{\text{DC}}+I_{\text{mod}}(t)$ and includes a constant intensity $I_{\text{DC}}$ and a harmonically modulated component
\begin{equation}
 I_{\text{mod}}(t)=I\exp(i\omega t)\quad\text{with}\quad\omega=2\pi\nu\,,\label{eq:I_mod}
\end{equation}
where $\nu$ is the modulation frequency and $I$ is the amplitude of the modulated intensity.\footnote{In fact, the modulated intensity $I_{\text{mod}}(t)$ corresponds only to the real part of the expression on the right-hand side of~\ref{eq:I_mod}. For simpler calculations, the complex representation is used in this paper.} Since only the alternating signal is considered in the analysis, the modulated part of the initial intensity $I_{\text{mod}}(t)$ is compared with the modulated part of the outgoing intensity\footnote{Whenever intensities are mentioned in the following, this refers only to the harmonically modulated part, including amplitude and phase. This does not fully represent the measured intensity, but since the constant component provides no further information, it is not considered in the calculations.}
\begin{equation}
 I_{\text{out}}(t)=\sum_{n=0}^{\infty}I_n(t)\label{eq:I_sum}
\end{equation}
leaving the cavity at time $t$ (cf.  Fig. \ref{fig:Resonator}). This total intensity is composed of an infinite number of individual rays $I_n(t)$, which have been reflected back and forth $n$ times and have therefore been attenuated to different degrees. The intensity of the $n$ times reflected ray can be described by
\begin{equation}
 I_n(t)=T^2\left(R^2\right)^{n}\exp\left(-\frac{\left(2n+1\right)L}{\Latt}\right)\cdot I_{\text{mod}}\left(t-\Delta t_n\right)\label{eq:I_n}
\end{equation}
and takes into account the attenuation by the sample according to Beers' Law (cf. Eq. \ref{eq:Beer}) as well as the reflection and transmission coefficients $R$ and $T=1-R$. The ray $I_n(t)$ has been reflected $n$ times inside the cavity and has entered the resonator at the time $t-\Delta t_n$, where $\Delta t_n=(n+1/2)\cdot\tau$ is considering the round trip time
\begin{equation}
 \tau=\frac{2L}{v_g}\label{eq:round_trip}
\end{equation}
of the cavity. The round trip time depends on the group velocity $v_g$ of the sample inside the cavity and the length $L$ of the resonator and describes the time required for the beam to complete one cycle in the resonator.

In order to compare the amplitude and the phase of the intensity leaving the cavity $I_{\text{out}}(t)$ with the initial intensity $I_{\text{mod}}(t)$, the infinite sum in~\ref{eq:I_sum} can be converted into the exponential representation of a harmonic function
\begin{equation}
 I_{\text{out}}(t)\equiv\tilde{I}(\nu)\exp\left(i\left[2\pi\nu t -\Delta\phi(\nu)\right]\right)\,,\label{eq:I_out}
\end{equation}
using an approach similar to~\cite{RingDown}, which is shown in the \ref{sec:I_out}. The frequency-dependent comparison of the reference and cavity beam intensities allows the attenuation length and the group velocity of the
sample to be precisely determined. This will be discussed in the following section.

\subsection{Frequency-dependent analysis of the intensities}
\label{sec:CELLPALS_amp_ratio_phase_shift}

\begin{figure}[htb]
	\centering
	\includegraphics[width=\picsize]{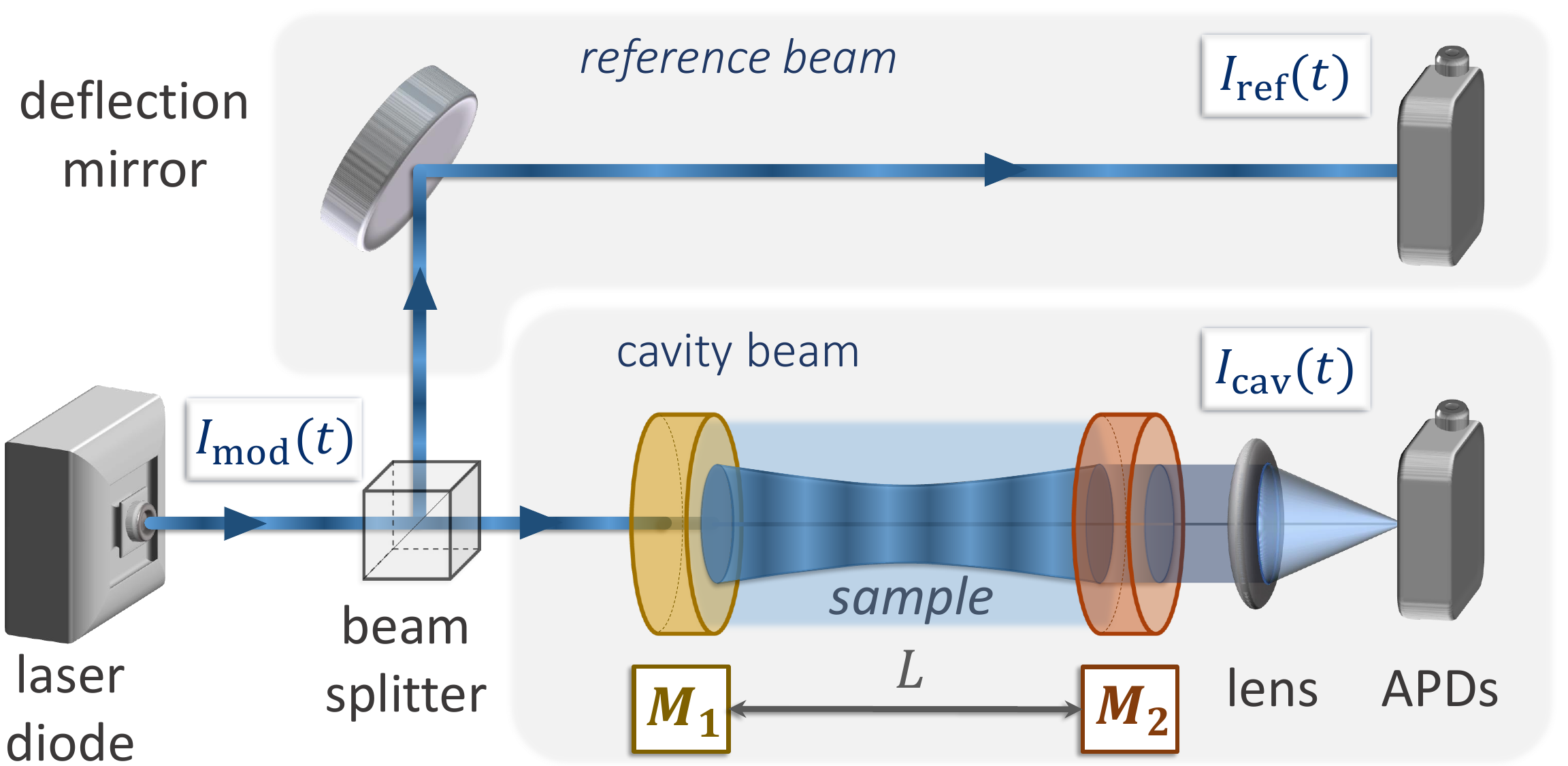}
	\caption{Simple schematic illustration of the \ac{CELLPALS} setup, showing the cavity resonator of length $L$ between the two mirrors $M_1$ and $M_2$ containing the sample. The initial modulated intensity $I_{\text{mod}}(t)$ of the laser diode is split into a reference beam and a cavity beam. The \acs{APD}s detect the respective frequency-dependent intensities $I_{\text{ref}}(t)$ and $I_{\text{cav}}(t)$.}
	\label{fig:Simp_setup}
\end{figure}

To compare the initial and the attenuated intensity, the laser beam is split into a reference beam with the intensity $I_{\text{ref}}(t)$ and a cavity beam with the intensity $I_{\text{cav}}(t)$, which are both detected by an \ac{APD} (cf.  Fig. \ref{fig:Simp_setup}). The modulated intensity of the reference beam
\begin{equation}
 I_{\text{ref}}(t)=I_{\text{mod}}(t)\,\xi_{\text{ref}}\exp(-i\phi_{\text{ref}}(\nu))\label{eq:I_ref}
\end{equation}
is equivalent to the modulated intensity $I_{\text{mod}}(t)$ (cf. Eq. \ref{eq:I_mod}) modulo an additional frequency-dependent phase $\phi_{\text{ref}}(\nu)=2\pi(x_{\text{ref}}/\vgair)\nu$ that occurs due to the path $x_{\text{ref}}$ from the beam splitter to the reference \acs{APD}, where $\vgair$ denotes the group velocity of light in air. 
The constant factor $\xi_{\text{ref}}$ represents all additional attenuation factors due to optical elements, such as the beam splitter or the deflection mirror (cf.  Fig. \ref{fig:Simp_setup}), which of course depend on the wavelength but do not change with different modulation frequencies $\nu$ of the intensity.

Similarly, the intensity of the cavity beam
\begin{equation}
 I_{\text{cav}}(t)=I_{\text{out}}(t)\,\xi_{\text{cav}}\exp(-i\phi_{\text{cav}}(\nu))\quad\label{eq:I_cav}
\end{equation}
corresponds to the intensity $I_{\text{out}}(t)$ (cf. Eq. \ref{eq:I_out}) modulo a linear phase $\phi_{\text{cav}}(\nu)=2\pi(x_{\text{cav}}/\vgair)\cdot\nu$ and the frequency-independent attenuating factor $\xi_{\text{cav}}$.

By measuring the intensities $I_{\text{ref}}(t)$ and $I_{\text{cav}}(t)$ for different modulation frequencies $\nu$, the amplitudes and the phases of both can be derived and compared to each other. On the one hand, this provides the amplitude ratio
\begin{equation}
 \amp(\nu)=\frac{\left\lvert  I_{\text{cav}}(t)\right\rvert  }{\left\lvert  I_{\text{ref}}(t)\right\rvert  }=\frac{b_1}{\sqrt{\Aeff^2-2\Aeff\cos\left(2\pi\frac{\nu}{\df}\right)+1}}\label{eq:A(nu)}
\end{equation}
for different modulation frequencies $\nu$, taking into account the  frequency-independent parameter $b_1=({\xi_{\text{cav}}}/{\xi_{\text{ref}}})\,T^2\exp(-{L}/{\Latt})$. On the other hand, it provides the frequency-dependent phase shift
\begin{align}
 \phase(\nu)&=\arg\left(I_{\text{ref}}(t)\right)-\arg\left(I_{\text{cav}}(t)\right)\notag\\[3pt]
 &=\arctan\left(\frac{\Aeff\sin\left(2\pi\frac{\nu}{\df}\right)}{1-\Aeff\cos\left(2\pi\frac{\nu}{\df}\right)}\right)+b_2\cdot\nu\label{eq:phi(nu)}
\end{align}
between the modulation of the cavity and the reference beam. The linear part of the phase shift $b_2\cdot\nu={\pi\tau\nu}+[\phi_{\text{cav}}(\nu)-\phi_{\text{ref}}(\nu)]$ reflects the effect of the path differences between the beam splitter and the corresponding \acs{APD}s and therefore does not contain any physical information about the sample. 
Both the amplitude ratio $\amp(\nu)$ and the phase shift $\phase(\nu)$ depend on the free spectral range
\begin{equation}
 \df=\frac{v_g}{2L}\,,\label{eq:delta_nu}
\end{equation}
which is defined by the ratio of the group velocity $v_g$ and the cavity length $L$ and corresponds to the inverse of the round trip time $\tau$ (cf. Eq. \ref{eq:round_trip}), and the attenuation parameter
\begin{equation}
 \Aeff=R^2\exp\left(-\frac{2L}{\Latt}\right)<1\,,\label{eq:A_eff}
\end{equation}
which takes into account the reflectivity $R$ of the mirrors and the attenuation length $\Latt$ of the sample.

By determining the free spectral range $\df$ and the attenuation parameter $\Aeff$ from the frequency-dependent measurements of the amplitude ratio $\amp(\nu)$ or the phase shift $\phase(\nu)$, the attenuation length $\Latt$ and the group velocity $v_g$ of the sample in the cavity can be derived.

\section{Measurement method and achievable precision}
\label{cap:precision}

In the \ac{CELLPALS} method, the intensity of the laser diode is sinusoidally modulated with modulation frequencies between $\sim\unit[0.5]{MHz}$ and $\sim\unit[350]{MHz}$ for a selected resonator length $L$, and the amplitudes and phases of the reference and sample beam are determined from intensity measurements for each modulation frequency $\nu$. In this way, the frequency-dependent amplitude ratio and phase shift between the two signals are derived, which can be fitted by the Eqs.~\ref{eq:A(nu)} and \ref{eq:phi(nu)} using the least squares method to estimate three parameters ($\df$, $\Aeff$ and $b_1$ respectively $b_2$) for both fit functions. 
\begin{figure}[htb]
	\centering
	\includegraphics[width=\picsize]{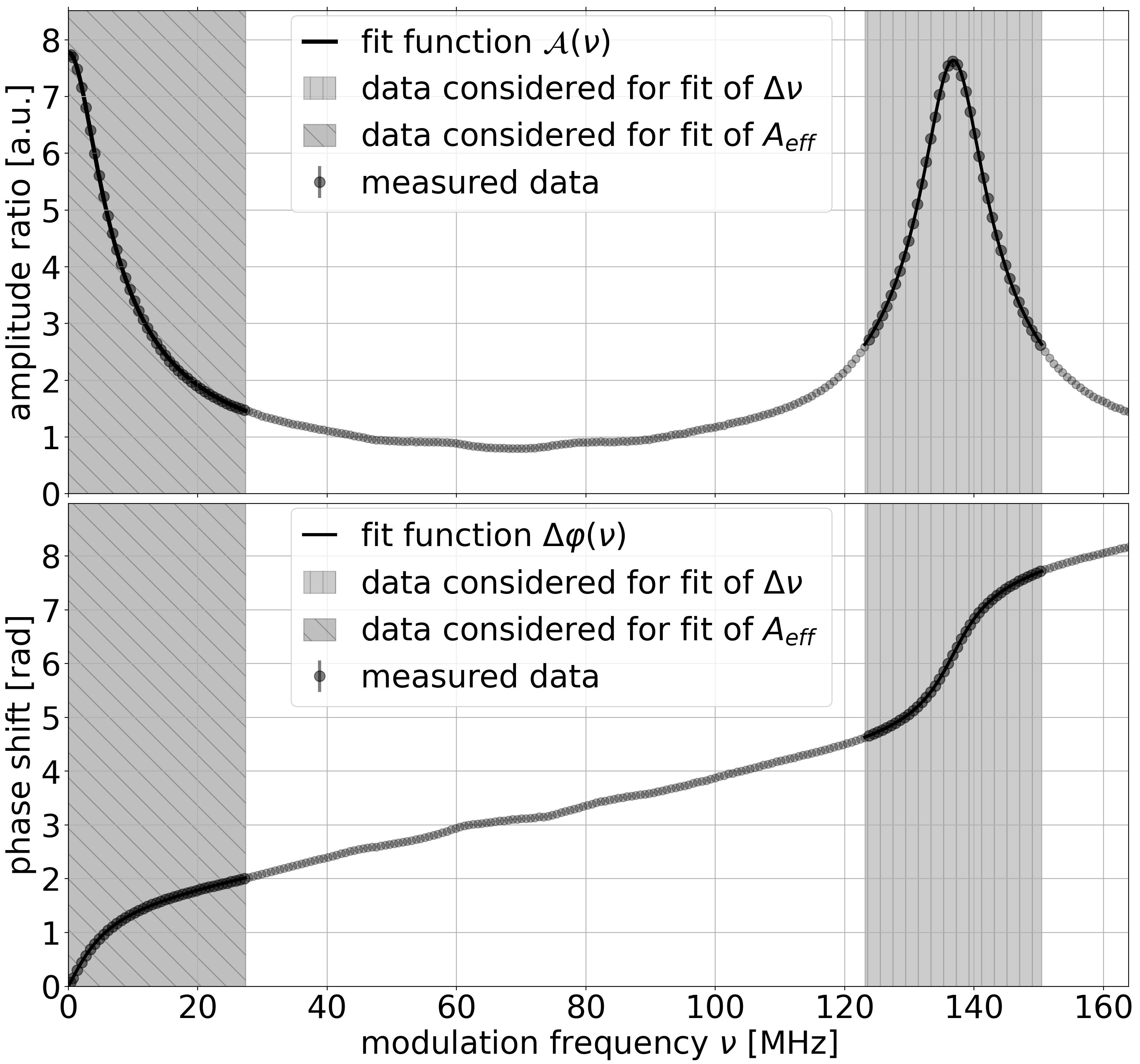}
	\caption{Measured amplitude ratio $\amp(\nu)$ (upper plot) and phase shift $\phase(\nu)$ (lower plot) shown exemplarily for a cavity of length $L=\unit[70]{cm}$ filled with \ac{LAB} (Eqs.~\ref{eq:A(nu)} and \ref{eq:phi(nu)} are used as fitting functions). The free spectral range $\df$ is determined using the least squares fitting method, taking into account the data in the marked area around the first order peak. The attenuation parameter $\Aeff$, on the other hand, is determined using the least squares fitting method, taking into account only the data in the falling edge of the zero order peak.}
	\label{fig:Data_example}
\end{figure}

\subsection{Uncertainty of the fit parameters} 
\label{sec:stat_fit_params} 

Since the fit parameters $\df$ and $\Aeff$ are used to determine the group velocity $v_g$ and the attenuation length $\Latt$ (cf. Eqs. \ref{eq:delta_nu} and \ref{eq:A_eff}), the uncertainties of these parameters have to be analysed. Fig. \ref{fig:Data_example} shows an exemplary measurement of a cavity of length $L=\unit[70]{cm}$ filled with \ac{LAB}. It has proven useful to restrict the fit range to modulation frequencies $\nu$ where the intensity is not too low.

The free spectral range is therefore determined in the range  $[0.9,1.1]\df$ (cf. vertically hatched area in  Fig. \ref{fig:Data_example}), as this parameter indicates the position of the peak $\nu=\df$ (constructive interference). The attenuation parameter $\Aeff$ is then determined in a second fit in the range $[0.0,0.2]\df$ using the estimated value for $\df$ (cf. diagonally hatched area in Fig. \ref{fig:Data_example}). In this range, the fit is in excellent agreement with the data for all measurements taken. 
\begin{figure}[htb]
	\centering
	\includegraphics[width=\picsize]{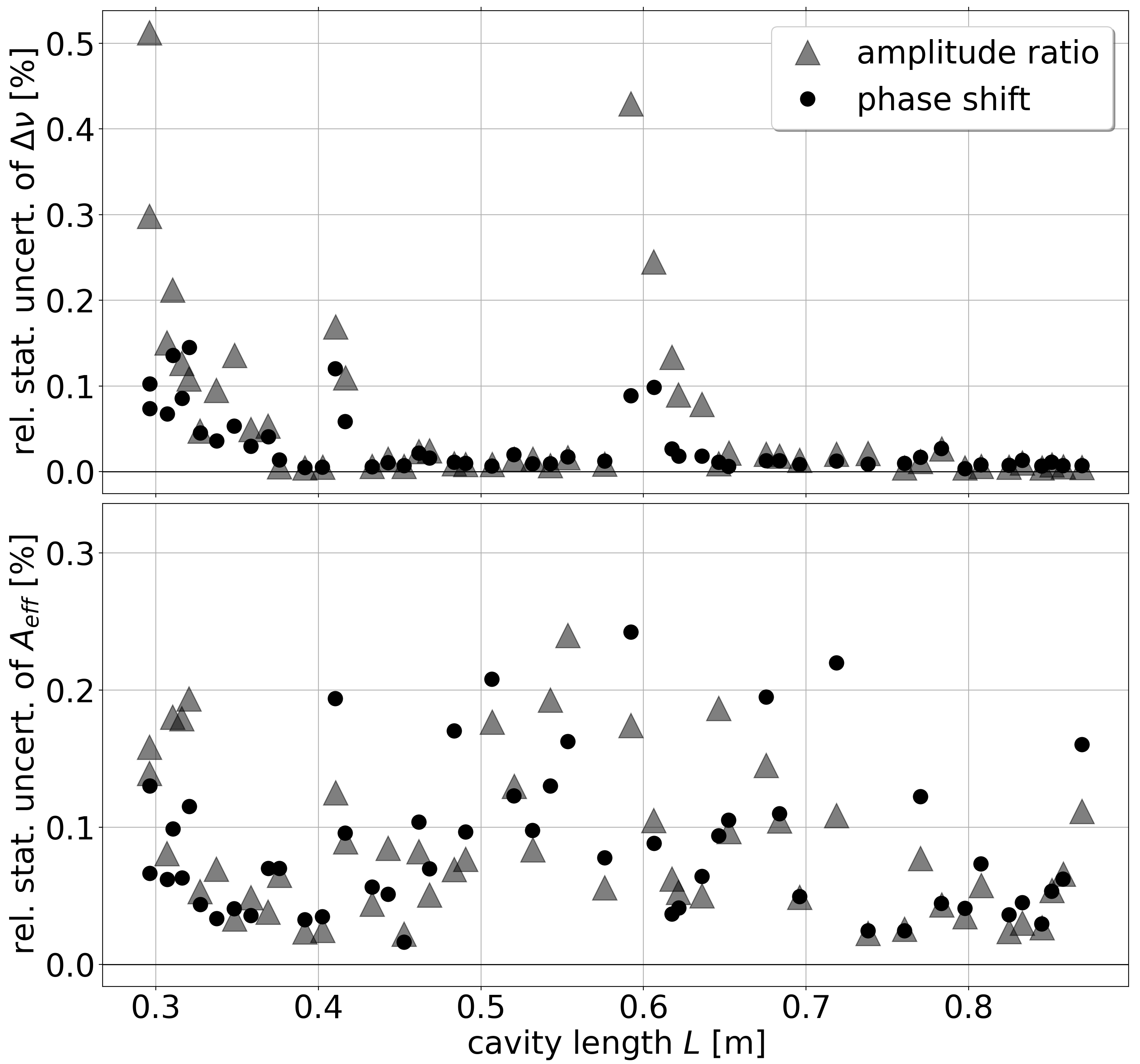}
	\caption{Relative uncertainty of the estimated fit parameters $\df$ and $\Aeff$ of a cavity filled with \ac{LAB}. The standard deviation was estimated from a total of 15,675 measurements at 51 different cavity lengths. The grey triangles show the relative uncertainties obtained from the amplitude ratio $\amp(\nu)$ and the black dots show the results obtained from the phase shift $\phase(\nu)$. The fit parameters $\df$ (upper plot) and $\Aeff$ (lower plot) were estimated by the least squares fitting method in the respective intervals described in figure Fig. \ref{fig:Data_example}.}
	\label{fig:Fit_param_sta_uncertainty}
\end{figure}

In order to estimate the magnitude of the fluctuations in the fit parameters, a total of 15,675 measurements were taken at 51 different cavity lengths ranging from $\unit[30]{cm}$ to $\unit[90]{cm}$ filled with \ac{LAB}. For each length, the mean and standard deviation of the estimated fit parameters $\df$ and $\Aeff$ were determined from both the amplitude ratio and the phase shift and shown as relative uncertainty in  Fig. \ref{fig:Fit_param_sta_uncertainty}. For each cavity length, the number of measured frequencies in the free spectral width was kept constant to ensure comparability of the fit uncertainty.

The uncertainty of the free spectral range $\df$ (cf. upper plot in  Fig. \ref{fig:Fit_param_sta_uncertainty}) increases for smaller cavity lengths, since for $\nu>\unit[350]{MHz}$ the harmonically modulated signal is distorted due to the limited bandwidth of the hardware used (cf.  \ref{sec:Exp_setup}). The higher uncertainties at $L=\unit[0.6]{m}$ and $L=\unit[0.3]{m}$ are related to the geometry of the beam, as the resonator mirrors have a radius of curvature of $r=\unit[0.6]{m}$. With an average relative uncertainty of $\sim\unit[0.03]{\%}$, the values derived from the phase shift (cf. black dots in the upper plot of  Fig. \ref{fig:Fit_param_sta_uncertainty}) are better suited for determining the free spectral range $\df$ than the data derived from the amplitude ratio (cf. grey triangles), whose average relative uncertainty is $\sim\unit[0.07]{\%}$.

In the same way, the relative uncertainties for the attenuation parameter $\Aeff$ are shown in the lower plot of  Fig. \ref{fig:Fit_param_sta_uncertainty}. On average, the uncertainty is $\sim\unit[0.09]{\%}$ for the data obtained from the amplitude ratio and from the phase shift. However, in some cases the amplitude ratio is preferred to the phase shift for estimating the attenuation parameter $\Aeff$, since the estimated value is less dependent on the respective fitting interval.

\subsection{Achievable precision of the group velocity and the cavity length} 
\label{sec:vel_uncert}

The group velocity $v_g$ in the sample can be determined with  \ac{CELLPALS} according to
\begin{equation}
    v_g=\frac{\df}{\df_{\text{air}}}\cdot\vgair\label{eq:v_g}
\end{equation}
by measuring the free spectral width in air $\df_{\text{air}}$ and in the sample $\df$ for a chosen resonator length $L$ (cf. Eq. \ref{eq:delta_nu}). The literature value of the group velocity in air
\begin{equation}
    \vgair^{\text{lit}}(\lambda=\unit[430]{nm})=\unit[(0.999700\pm 0.000005)]{c}\label{eq:v_g_air_lit}
\end{equation}
is known very precisely and can be derived from the chromatic dispersion ($c$ is the speed of light in vacuum)~\cite{Ciddor}. Since this value is known about 60 times more precisely than the relative uncertainty
of the free spectral width $\df$ (cf.  \ref{sec:stat_fit_params}), the achievable precision of \ac{CELLPALS} to measure the group velocity $v_g$ in a sample is limited by this uncertainty of $\sim\unit[0.03]{\%}$.  

On the other hand, according to~\ref{eq:delta_nu}, the cavity length
\begin{equation}
    L=\frac{\vgair^{\text{lit}}}{2\cdot\df_{\text{air}}}\label{eq:length_cav}
\end{equation}
can also be determined in air with the same accuracy of $\unit[0.03]{\%}$ if the group velocity is known.
\begin{figure}[htb]
	\centering
	\includegraphics[width=\picsize]{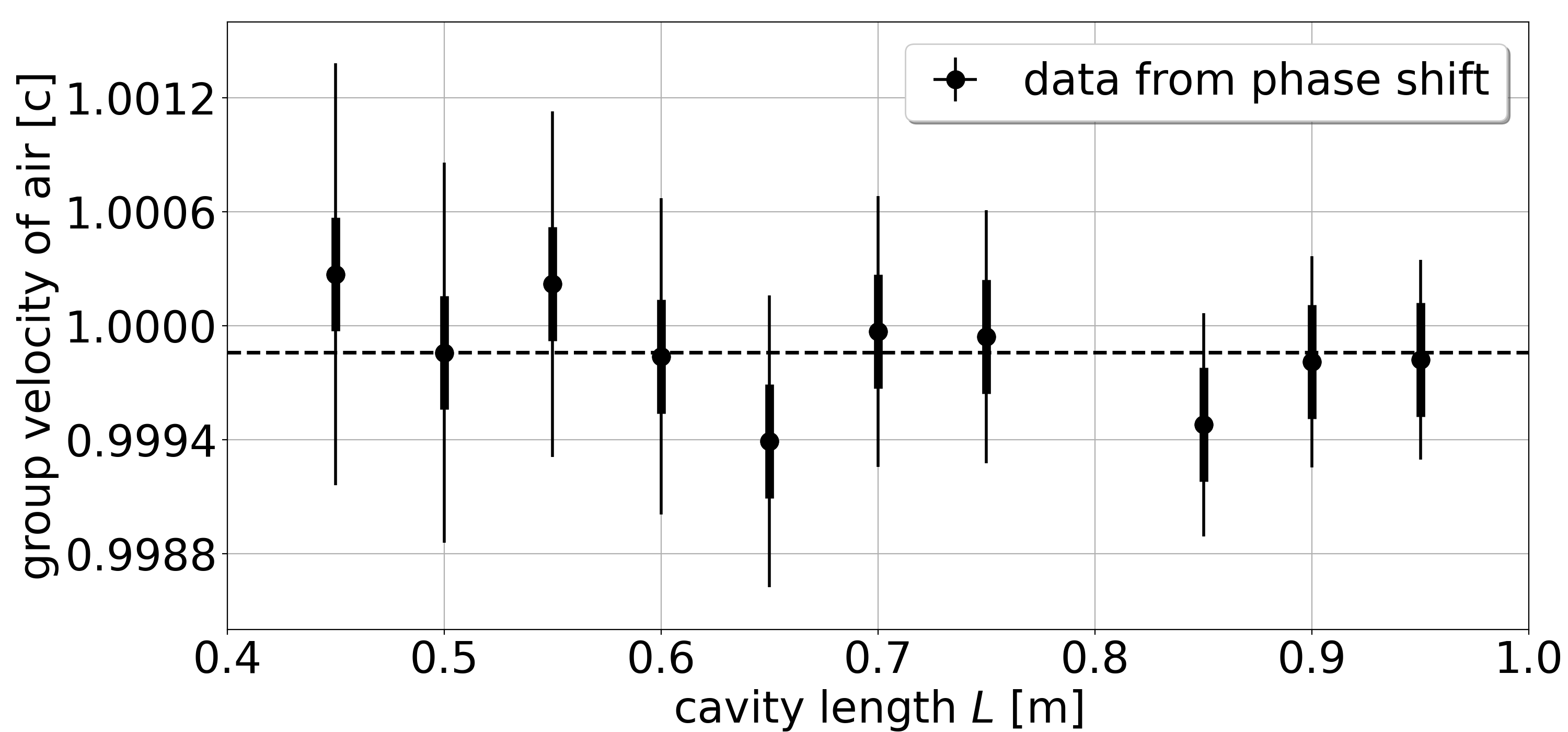}
	\caption{Measurement of the group velocity in air with empty resonators. The resonator length $L$ was measured with the yardstick with an estimated uncertainty of $\unit[0.5]{mm}$, which corresponds to the thin errorbars (cf. Eq. \ref{eq:delta_nu}). The data were derived from the fitted free spectral range $\df$ and the corresponding uncertainties (indicated by the thicker errorbars) of the phase shift. The mean value of the group velocity in air at $\lambda=\unit[430]{nm}$ was determined to be $v_{g,\text{air}}=\unit[(0.99986\pm 0.00083)]{c}$.}
	\label{fig:v_g_air}
\end{figure}
To check whether $\df$ is indeed suitable for determining the group velocity, a measurement with air was performed at several cavity lengths (cf.  Fig. \ref{fig:v_g_air}) and compared to the literature value in~\ref{eq:v_g_air_lit}. For this measurement the lengths of the cavities were determined with a yardstick and estimated with an uncertainty of $\unit[0.5]{mm}$, which according to~\ref{eq:delta_nu} leads to an uncertainty of the derived group velocity (cf. thin error bars in  Fig. \ref{fig:v_g_air}). For each cavity length, the group velocity was derived from the free spectral range of the phase shift, considering an uncertainty of $\unit[0.03]{\%}$. Taking into account the quadratic error propagation of both uncertainties, the group velocity in air was determined to be $v_{g,\text{air}}=\unit[(0.99986\pm 0.00083)]{c}$. Within the error bars, this result is consistent with the literature value in~\ref{eq:v_g_air_lit}, which demonstrates the suitability of the \ac{CELLPALS} method for determining the group velocity.

\subsection{Achievable precision of the attenuation length}
\label{sec:Latt_uncert}

In order to determine the attenuation length $\Latt$ with only one cavity, the reflectivity $R$ of the mirrors must be known. The manufacturer specifies the reflectivity in air as $R=0.95\pm0.01$, but this value
changes in contact with different liquids~\cite{BeamOptic}. If this uncertainty is taken into account for a cavity length of $L=\unit[0.5]{m}$, the systematic uncertainty $\text{d}\Latt^{\text{syst}}/\Latt=\left\lvert  \Latt L^{-1}\cdot \text{d}R^{\text{syst}}R^{-1}\right\rvert  $ for media with an attenuation length
$\Latt>\unit[10]{m}$ is more than $\unit[20]{\%}$. The uncertainty of the reflectivity provides the largest systematic error when determining the attenuation length with only one cavity length, therefore it should be eliminated. This
can be achieved by determining the attenuation parameter $\Aeff(L)$ for more than one resonator length. A variable length cavity design is used in \ac{CELLPALS} (cf.  \ref{sec:Exp_setup}), which provides any resonator
length $L$ on a continuous scale. The maximum length $L_{\text{max}}=2r$ is limited by the stability criterion in~\ref{eq:stability_criterion}, which depends on the radii of curvature $r$ of the concave
resonator mirrors. The minimum length $L_{\text{min}}=v_g/(2\df_{\text{max}})$ corresponds to the maximal detectable frequency of the free spectral range $\df_{\text{max}}\sim\unit[350]{MHz}$ which is limited by the hardware components of the setup and depends on the group
velocity of the sample (cf.  \ref{sec:Exp_setup}).

In order to determine $\Aeff(L)$, however, the chosen cavity length $L$ must be known for each measurement. For this purpose, a reference length $L_0$ is determined according to~\ref{eq:length_cav} by measuring the free spectral width in air $\df_{0,\text{air}}$. After that, the sample is filled into the cavity and the new free spectral width $\df_0$ is measured at the same resonator length $L_0$. Any new length $L_i=\df_0/\df_i\cdot L_0$ chosen can now be determined by the new free spectral width $\df_i$. The attenuation length $\Latt$ is determined from the measured attenuation parameters $\Aeff(L)$, using the least squares fitting method with Eq. \ref{eq:A_eff} as the fit function.

\begin{figure}[htb]
	\centering
	\includegraphics[width=\picsize]{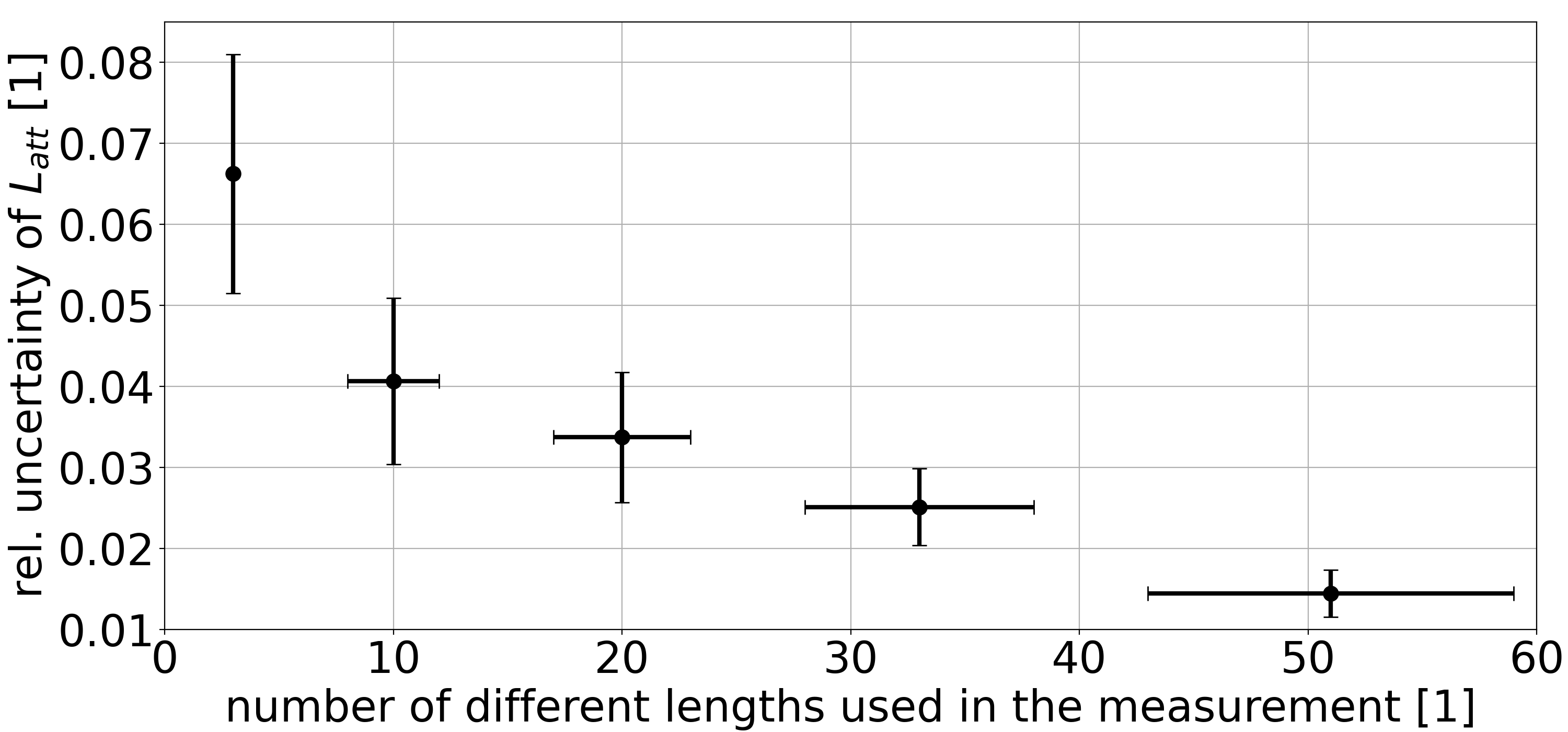}
	\caption{Estimation of the relative uncertainty of the derived attenuation length $\Latt$ as a function of the number of different cavity lengths that have been used in the measurement. If more than 50 different lengths are measured, the relative uncertainty can be reduced to less than $\unit[2]{\%}$.}
	\label{fig:uncert_VLC}
\end{figure}

The achievable precision of the attenuation length measurement depends on the number of resonator lengths used, as demonstrated in  Fig. \ref{fig:uncert_VLC}. With only three different lengths, a relative uncertainty of about $\unit[7]{\%}$ can be achieved, which, however, can be improved to less than $\unit[2]{\%}$ if more than 50 data points are recorded.

With this measurement method, the systematic uncertainty due to the lack of precise knowledge of the mirror reflectivity $R$ is eliminated. This method can even be used to determine the reflectivity of the mirrors in different media with a higher precision than the manufacturer's specifications.

\subsection{Systematic effects of incomplete collimation of the cavity beam}
\label{sec:Uncert_collimation}

Due to the beam cross-section of the laser light and the small sensitive area of the \acs{APD} (cf.  \ref{sec:Exp_setup}), it has to be ensured that the total intensity of the cavity beam is detected. In order to estimate
the effect of incomplete collimation on the measurements, the expected beam alignment for a certain cavity length was calculated by using Gaussian beam optics and ray transfer matrix~\cite{BeamOptic}.

\begin{figure}[hbt]
	\centering
	\includegraphics[width=\picsize]{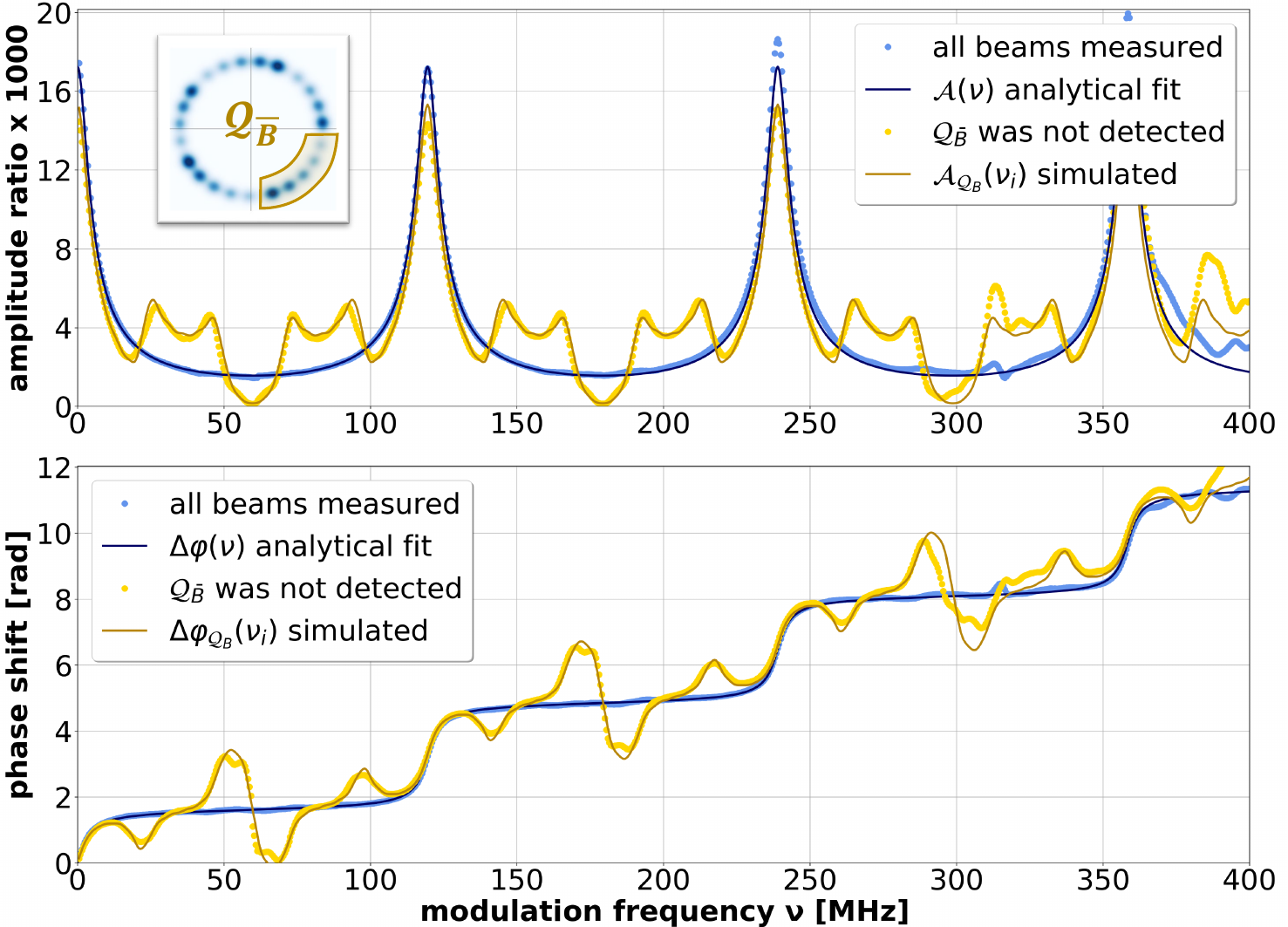}
	\caption{The blue dots show the measured signal of a $L=\unit[80]{cm}$ cavity filled with \ac{LAB} when all the rays were detected. The yellow line shows the calculated signal when the beams in $\mathcal{Q}_{\tilde{B}}$ were not detected by the \ac{APD}. The yellow dots show the corresponding measurements where the $\mathcal{Q}_{\tilde{B}}$ part of the beam alignment was covered with a sheet of paper. For modulation frequencies $<\unit[300]{MHz}$ the measured data are in excellent agreement with the calculations.}
	\label{fig:Sim_A}
\end{figure}
As the cavity meets the stability criterion
\begin{equation}
    0\leq g^2\leq 1\quad\text{with}\quad g=1+\frac{L}{r}\,,\label{eq:stability_criterion}
\end{equation}
which depends on the cavity length $L$ and on the radii of curvature $r$ of the concave resonator mirrors, the rays do not leave the cavity. If the outgoing intensity is not fully collimated, there will be an infinite number of individual rays $I_n(t)$ which, according to~\ref{eq:I_out}, no longer contribute to the total outgoing intensity $I_{\text{out}}(t)$. 

In  Fig. \ref{fig:Sim_A}, the blue dots show the fully collimated measurements of the amplitude ratio (upper plot) and the phase shift (lower plot) of an $\unit[80]{cm}$ cavity filled with \ac{LAB}. In this case, the data is in agreement with the analytical fit functions in Eqs.~\ref{eq:A(nu)} and \ref{eq:phi(nu)}, which is not the case for incomplete collimation (yellow dots). To estimate the effect of incomplete collimation, the expected signal for the same cavity was calculated if the rays indicated by $\mathcal{Q}_{\tilde{B}}$ (cf. calculated beam alignment in the upper left subplot in  Fig. \ref{fig:Sim_A}) do not contribute to the total intensity (cf. yellow line). This result was experimentally confirmed by covering the respective rays in $\mathcal{Q}_{\tilde{B}}$ with a piece of paper (cf. yellow dots). Several analyses of this kind show that an incomplete collimation of the outgoing intensity always results in characteristic deviations, which occur correlated in the amplitude ratio and in the phase shift and are also reflected in some measurements. These characteristic deviations provide a strong but simple criterion for incomplete collimation and allow us to dismiss such measurements.

\section{Experimental setup of CELLPALS}
\label{sec:Exp_setup}

In this chapter, the experimental setup for the measurements with \ac{CELLPALS} is presented. The required hardware components are discussed in more detail in  \ref{sec:Hardware} and the optical setup, including the variable-length cavity, is introduced in  \ref{sec:OptSetup}.

\subsection{Hardware and electronics}
\label{sec:Hardware}

\begin{figure}[htb]
	\centering
	\includegraphics[width=\picsize]{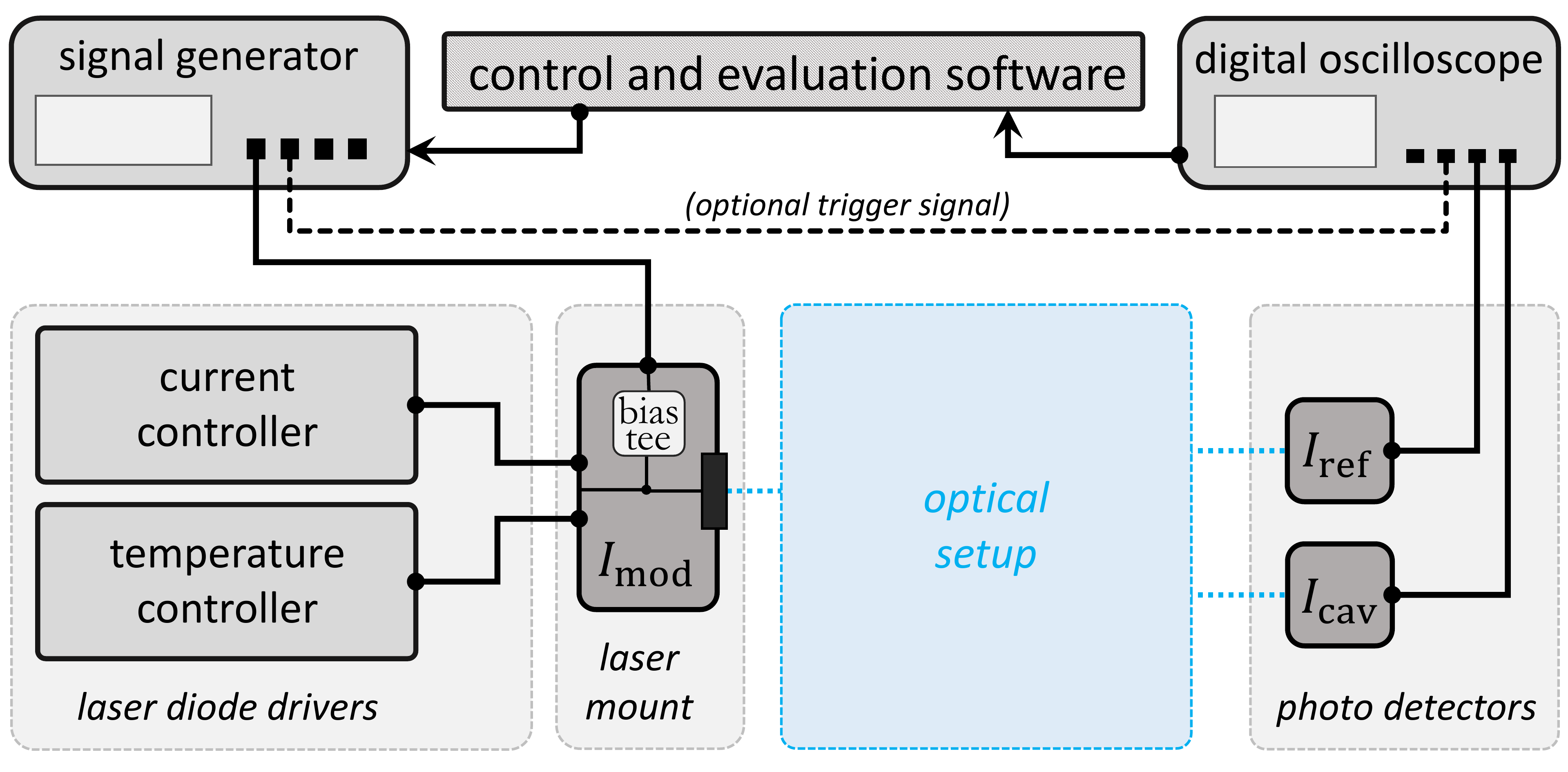}
	\caption{Schematic illustration of the hardware components and the control and measurement software of \ac{CELLPALS}. The optical setup is shown in figure Fig. \ref{fig:Simp_setup}.}
	\label{fig:Hardware}
\end{figure}

The experimental setup including the hardware devices for signal generation and readout electronics (cf. Fig. \ref{fig:Hardware}) is located in a dark box to avoid external light. The \ac{LDs} of the type Roithner RLT4XX-50CMG with wavelengths $\unit[420]{nm}$, $\unit[425]{nm}$, $\unit[430]{nm}$ and $\unit[435]{nm}$ used in the measurements in  \ref{sec:Measurements} are specified with a FWHM of $<\unit[2]{nm}$ for the peak wavelength~\cite{LD}. The
\ac{LDs} are mounted in direct thermal contact with the cold plate in the laser mount~\cite{LaserMount}, which is cooled to a constant temperature of $\unit[20]{\text{\textdegree}C}$ by the temperature controller~\cite{LaserDriverTemp}. The
current controller supplies a constant current $I_{\text{DC}}$~\cite{LaserDriverCurrent}, which was set to a value larger than the specified threshold current of the corresponding \ac{LDs}~\cite{LD}.

A modulation signal $V_{\text{mod}}(t)$ is provided by a signal generator of the type DSG815 with the bandwidth $\unit[350]{MHz}$~\cite{FG}. Inside the laser mount, the modulated voltage is converted into a current $I_{\text{mod}}(t)={V_{\text{mod}}(t)}/{Z_{\text{in}}}$ via a
bias tee (considering the impedance $Z_{\text{in}}$), which adds the converted modulated current $I_{\text{mod}}(t)$ to the constant current $I_{\text{DC}}$ provided by the controller. Frequencies $\nu$ of the modulated signal
between $\unit[100]{kHz}$ and $\unit[500]{MHz}$ are supported by the laser mount~\cite{LaserMount}. Since the optical output power of the \ac{LDs} is approximately linear to the forward current~\cite{LD}, the laser intensity
includes a constant component and a sinusoidally modulated component.

After passing through the optical elements (cf.  Fig. \ref{fig:Simp_setup}), the light is focused on the sensitive area of $\unit[0.5\times0.5]{mm}^2$ of the \ac{APD}s (type APD430A/M)~\cite{APD}. 
The type DSO-X4034~A digital oscilloscope with the bandwidth $\unit[350]{MHz}$ and a maximum sampling rate of $\unitfrac[5]{GSa}{s}$~\cite{DO} is used to digitise the signals from the \ac{APD}s, optionally receiving an additional trigger
signal from the signal generator (indicated by the dashed line in  Fig. \ref{fig:Hardware}). The control and evaluation software is based on a user-specific Python script and ensures that the same number of data points is
measured for each chosen resonator length.

\subsection{Optical design and the variable-length cavity design}
\label{sec:OptSetup}

The optical design used in \ac{CELLPALS} is illustrated in  Fig. \ref{fig:Simp_setup}. The light emitted by the \ac{LDs} is initially split into a reference beam with $\unit[10]{\%}$ and a cavity beam with $\unit[90]{\%}$ of the initial intensity. For both beams, possible reflections are kept from falling back into the \ac{LDs} by optical isolators. Before being detected by the \ac{APD}s, the beams are collimated with a collecting lens so that the entire signal is detected and deviations due to incomplete collimation are prevented. The \ac{APD} detecting the cavity beam is aligned at an angle to the optical plane to prevent the effect of a second short resonator between the exit mirror of the cavity and the \ac{APD}.

\begin{figure}[htb]
	\centering
	\includegraphics[width=\picsize]{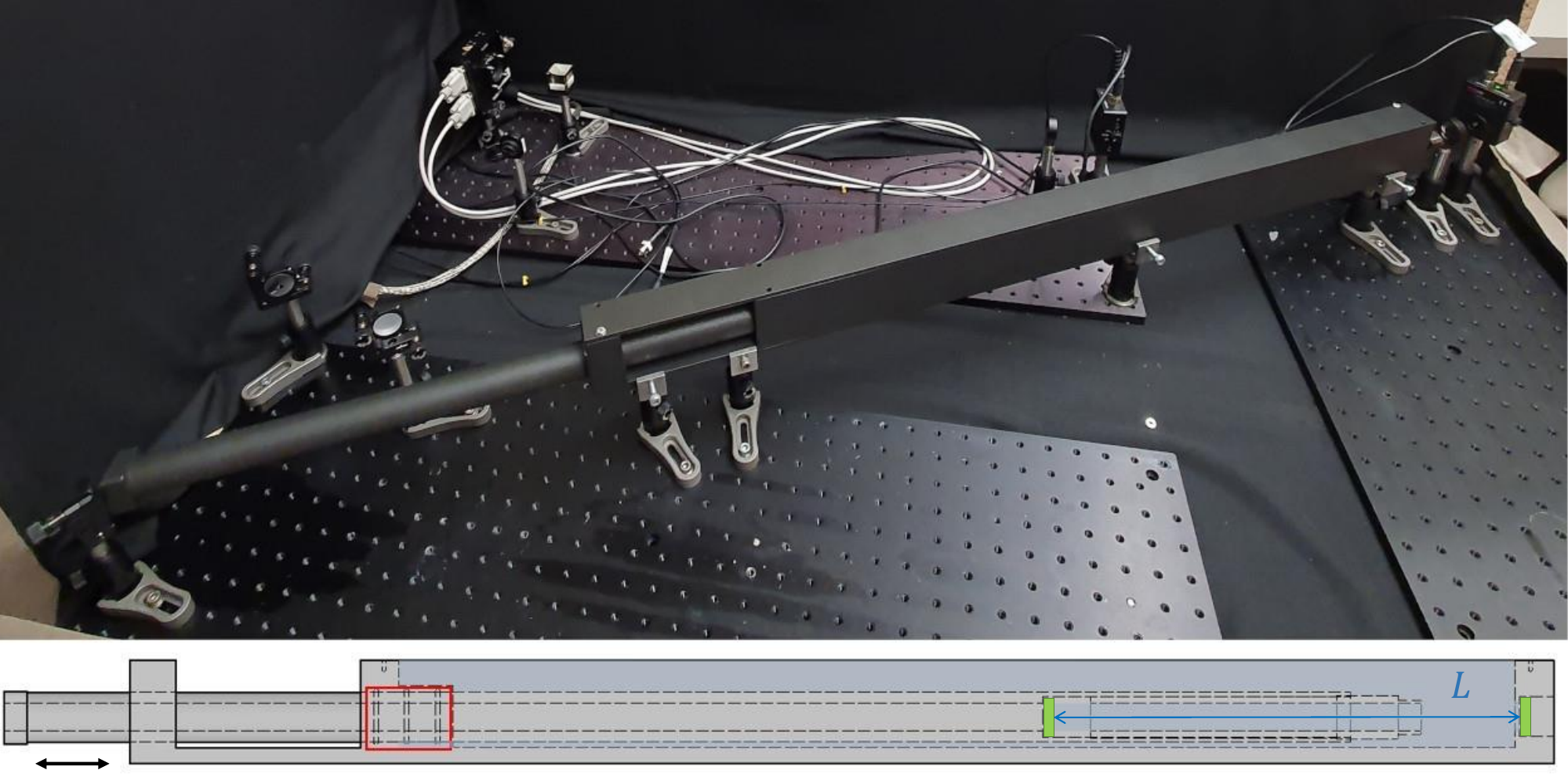}
	\caption{A picture of the optical setup (top) and a schematic illustration of the variable length cavity design (bottom). The entire cavity was made of stainless steel and then completely coated with \ac{PTFE}. It consists of a rectangular sample holder and a movable tube containing one of the two semi-transparent resonator mirrors (marked in green in the lower plot). The second mirror, through which the beam leaves the resonator, is fixed at the rear end of the frame. Three sealing rings (marked in red in the lower plot) ensure that the cavity is sealed.}
	\label{fig:VLC_pic}
\end{figure}

A variable-length cavity was designed, made of stainless steel and coated with adcoat\textregistered32/2889 \ac{PTFE} by the Adelhelm company to avoid possible reactions between the stainless steel and the sample (cf.  Fig. \ref{fig:VLC_pic}).
The variable-length cavity consists of a rectangular sample holder and a movable tube, where the length $L$ of the resonator is defined by the reflecting surface of the two opposing semi-transparent mirrors and can be continuously varied from $\sim\unit[25]{cm}$ to $\sim\unit[90]{cm}$. One of the mirrors is fixed at the rear end of the sample holder, whereas the other is mounted in the movable tube (cf. green marked areas in the lower plot in  Fig. \ref{fig:VLC_pic}). The concave mirrors were manufactured by LensOptics with a reflectivity of $R=0.95\pm0.01$ in the wavelength range $\unit[(420-430)]{nm}$, a diameter of $1"$ and a radius of curvature of $r=\unit[0.6]{m}$. The tube is sealed at the point of contact with the cavity using three \ac{PTFE} sealing rings, similar to a dynamic rod seal (cf. red marked area in  Fig. \ref{fig:VLC_pic}). This allows the pipe to be moved without liquid leaking out of the cavity. In this way, many measurements at different lengths can be performed without the need to empty and refill the cavity each time.

\section{UV-Vis method for determining the attenuation length} 
\label{sec:UV_Vis} 

A common method for determining the transparency of liquids is UV--Vis spectroscopy, which uses a cuvette filled with the sample. The attenuation of the intensity caused by the sample is analysed by comparing the original
intensity $I_0$ with the intensity $I(L)$ after traversing the cuvette of length $L$. Unlike \ac{CELLPALS}, however, the UV--Vis method does not use an optical resonator, so the effective light path
is limited to $L=\unit[10]{cm}$ with conventional devices. The UV--Vis method is based on a wavelength-dependent measurement of the absorbance $\Abs(L)$~\cite{UV-VIS} which can be transferred into the attenuation length
$\Latt(\Abs)$:
\begin{equation}
    \Abs(L)=\log_{10}\left(\frac{I_0}{I(L)}\right)\quad\Leftrightarrow\quad\Latt(\Abs)=\frac{L\cdot\log_{10}(e)}{\Abs{}(L)}\,.\label{eq:Latt_abs}
\end{equation}
The absorbance of a cuvette filled with a sample can be described by
\begin{equation}
    \Abs_{\text{full}}(\lambda)=2\Abs_{R_{CA}}(\lambda)+2\Abs_{R_{CS}}(\lambda)+2\Abs_{\text{w}}(\lambda)+\Abs_S(\lambda)\,,\label{eq:A_full}
\end{equation}
where
\begin{itemize}
    \item $\Abs_{R_{CA}}(\lambda)=-\log_{10}(1-R_{CA}(\lambda))$ describes the reflection at the surface between cuvette and air with the reflection coefficient $R_{CA}(\lambda)$
    \item $\Abs_{R_{CS}}(\lambda)=-\log_{10}(1-R_{CS}(\lambda))$ describes the reflection at the surface between the cuvette and the sample
    \item $\Abs_{\text{w}}(\lambda)=d_C/L_{\text{att,C}}(\lambda)\cdot\log_{10}(e)$ describes the attenuation by passing the glass of the cuvette
    \item $\Abs_S(\lambda)=L/\Latt(\lambda)\cdot\log_{10}(e)$ describes the absorbance by the sample and can be used to determine the attenuation length $\Latt(\lambda)$.
\end{itemize}
Since $\Abs_{R_{CA}}(\unit[430]{nm})\approx1.6\cdot 10^{-2}$~\cite{Ciddor1,FusedSilica} is significantly larger than the attenuation by the sample $\Abs_S(\Latt>\unit[10]{m})<4.3\cdot10^{-3}$, the absorbance of the empty cuvette
\begin{equation}
    \Abs_{\text{empty}}(\lambda)=4\Abs_{R_{CA}}(\lambda)+2\Abs_{\text{w}}(\lambda)\label{eq:A_empty}
\end{equation}
is also measured in the UV--Vis method. With
\begin{equation}
    \Abs_{\text{sub}}(\lambda)=\Abs_{\text{full}}(\lambda)-\frac{1}{2}\Abs_{\text{empty}}(\lambda)\geq0\,,\label{eq:A_obs}
\end{equation}
$\Abs_{R_{CA}}(\lambda)$ can be eliminated. According to Eqs.~\ref{eq:A_full}, \ref{eq:A_empty} and \ref{eq:A_obs}, the absorbance of the sample can now be determined by
\begin{equation}
    \Abs_S(\lambda)=\Abs_{\text{sub}}(\lambda)-\Abs_{R_{CS}}(\lambda)-\Abs_{\text{w}}(\lambda)\,.\label{eq:A_S}
\end{equation}
Since $\Abs_{R_{CS}}(\lambda)$ and $\Abs_{\text{w}}(\lambda)$ are not known exactly, the following assumptions are made: first, $\Abs_{R_{CS}}$ and $\Abs_{\text{w}}$ are assumed to be wavelength-independent in the measured range and, furthermore, an infinite transparency $\Abs_S(\lambda_{\text{min}})=0$ of the sample is assumed for the minimum absorbance at the wavelength $\lambda_{\text{min}}$. Thus $\Abs_{\text{min}}=\Abs_{\text{sub}}(\lambda_{\text{min}})$ is associated with $\Abs_{\text{min}}=\Abs_{R_{CS}}+\Abs_{\text{w}}$ and the absorbance of the sample is determined by
\begin{equation}
    \Abs_S(\lambda)=\Abs_{\text{sub}}(\lambda)-\Abs_{\text{min}}\label{eq:A_S_corr}
\end{equation}
and is then converted to the attenuation length according to~\ref{eq:Latt_abs}.
\begin{figure}[htb]
	\centering
	\includegraphics[width=\picsize]{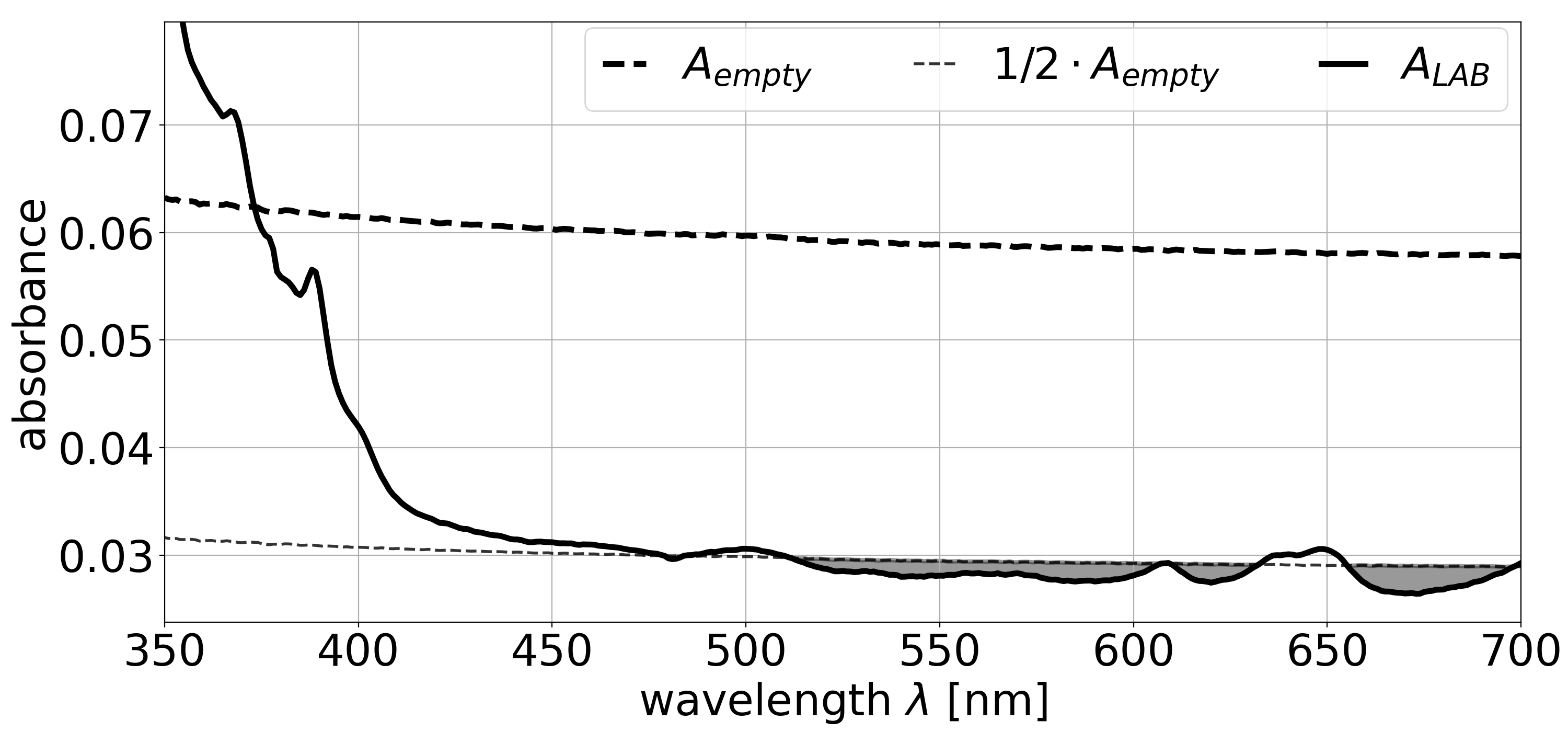}
	\caption{Absorbances $A_\text{full}(\lambda)$ of a \ac{LAB} sample and $A_\text{empty}(\lambda)$ of the empty cuvette measured with the PerkinElmer Lambda 850+ UV-Vis spectrophotometer. Also $1/2\cdot A_\text{empty}(\lambda)$ is added to the plot to emphasize the grey hatched area, where $\Abs_{\text{sub}}(\lambda)<0$ gives negative values, contrary to the theoretical considerations (cf. Eq. \ref{eq:A_obs}).}
	\label{fig:A_obs_LAB_ex}
\end{figure}

 Fig. \ref{fig:A_obs_LAB_ex} shows the measured absorbance of the full and empty cuvette $\Abs_{\text{full}}(\lambda)$ and $\Abs_{\text{empty}}(\lambda)$ for a typical sample. In addition, also \mbox{$1/2\cdot\Abs_{\text{empty}}(\lambda)$} is inserted into the plot, which illustrates that the subtracted absorbance $\Abs_{\text{sub}}$ in~\ref{eq:A_obs}, contrary to the theoretical consideration, provides negative values for the minimum absorbance. By setting this minimum value to $\Abs_{\text{sub}}(\lambda_{\text{min}})=0$, the absorbance is shifted towards higher values, which leads to a systematic underestimation of the attenuation length for highly transparent media. Since this effect cannot be explained by the theoretical considerations, it is difficult to estimate the extent of the resulting systematic uncertainty.

\section{Measurements and results}
\label{sec:Measurements}

In this chapter, selected results of the \ac{CELLPALS} method are presented. In  \ref{subsec:performance_measurement} the measurements of five samples with different attenuation lengths between $\sim\unit[1]{m}$ and $\sim\unit[40]{m}$ were analysed, and the results are compared to the UV--Vis results.  \ref{subsec:high_precision} shows a high-precision measurement of \ac{LAB} at two wavelengths ($\unit[420]{nm}$ and $\unit[430]{nm}$) and in  \ref{subsec:LAB_measurements}, the results of three ultra-pure samples from an optical purification plant for a \ac{LAB}-based \ac{LS} are presented. 

\subsection{Measurements of samples with varying degrees of transparencies with CELLPALS and the UV-Vis method}
\label{subsec:performance_measurement}

In order to compare the performance of the \ac{CELLPALS} method for different levels of transparency with the conventional UV--Vis method, five samples were measured with both \ac{CELLPALS} and a PerkinElmer 850+ spectrophotometer. The first sample EJ-309 is a xylene-based liquid scintillator manufactured by EljenTechnology and specified with $\Latt>\unit[1]{m}$. A \ac{LAB} sample manufactured by Helm AG was analysed, as well as a xylene sample from FisherScientific and a cyclohexane and pentane sample from Sigma Aldrich. The measured values of the attenuation length $\Latt$ are shown in 
Tab. \ref{tab:FiveSamples_Latt}.

\begin{table}[hbt]
		\centering
        \scalebox{0.95}{
		\begin{tabular}{|c|c||c||c|}
			\hline
            \multicolumn{2}{|c||}{\rule[-1.5ex]{0pt}{4.5ex} MEASUREMENT}&CELLPALS&\multicolumn{1}{c|}{\rule[-1.5ex]{0pt}{4.5ex} UV-Vis}\\
			\hline
			\rule{0pt}{15pt}sample & $\lambda$ [nm] &$\Latt$ [m]&$\Latt$ [m]\\ 
			\hline\hline 
			\rule{0pt}{15pt}EJ-309 & $435$ & $1.18\pm0.08$&$1.26\pm0.03$\\
			\hline
			\rule{0pt}{15pt}\ac{LAB} & $430$ &$9.02\pm0.66$&$8.43\pm1.19$\\ 
			\hline
			\rule{0pt}{15pt}xylene & $425$ &$9.67\pm0.38$&$7.56\pm0.97$\\ 
			\hline
			\rule{0pt}{15pt}cyclohexane & $425$ &$26.71\pm2.24$&$22.24\pm7.96$\\ 
			\hline
			\rule{0pt}{15pt}pentane & $435$&$37.36\pm2.51$&$22.67\pm12.47$\\ 
			\hline
		\end{tabular}}
        \caption{Measurements of the attenuation lengths $\Latt$ for different media derived with the \ac{CELLPALS} method measuring at three resonator lengths and the UV-Vis method. The measurements were taken at slightly different wavelengths $\lambda$ with an uncertainty of $<\unit[1]{nm}$ for \ac{CELLPALS} and $\sim\unit[0.02]{nm}$ for the UV-Vis method.}
		\label{tab:FiveSamples_Latt}
\end{table}
The attenuation lengths from the two methods are in agreement within their uncertainties. For attenuation lengths of less than $\sim\unit[10]{m}$, the \ac{CELLPALS} and UV--Vis methods show a similar uncertainty, although the results with only three resonator lengths do not reflect the maximum performance of the \ac{CELLPALS} method (cf.  \ref{sec:Latt_uncert}). However, for highly transparent media, the uncertainties obtained with the \ac{CELLPALS} method are significantly smaller.

\begin{table}[hbt]
		\centering
		\begin{tabular}{|c|c||c|}
			\hline
			\rule{0pt}{15pt}sample&$\lambda$ [nm] &$v_g/c$ [\%]\\ 
			\hline\hline 
			\rule{0pt}{15pt}EJ-309&$435\pm1$&$56.55\pm0.07$\\
			\hline
			\rule{0pt}{15pt}\ac{LAB}&$430\pm1$&$63.63\pm0.01$\\ 
			\hline
			\rule{0pt}{15pt}xylene&$425\pm1$&$61.55\pm0.02$\\ 
			\hline
			\rule{0pt}{15pt}cyclohexane&$425\pm1$&$67.49\pm0.01$\\ 
			\hline
			\rule{0pt}{15pt}pentane&$425\pm1$&$71.23\pm0.02$\\ 
			\hline
		\end{tabular}	
		\caption{Results of the group velocity $v_g$ for different media derived with the \ac{CELLPALS} method.}
		\label{tab:FiveSamples_v_g}
\end{table}
In addition to the attenuation length, the \ac{CELLPALS} method also provides the group velocity $v_g$ in the sample. 
Tab. \ref{tab:FiveSamples_v_g} shows the results for the five samples in units of the speed of light in a vacuum.

\subsection{High precision measurement of LAB with CELLPALS}
\label{subsec:high_precision}

\begin{figure}[htb]
	\centering
	\includegraphics[width=\picsize]{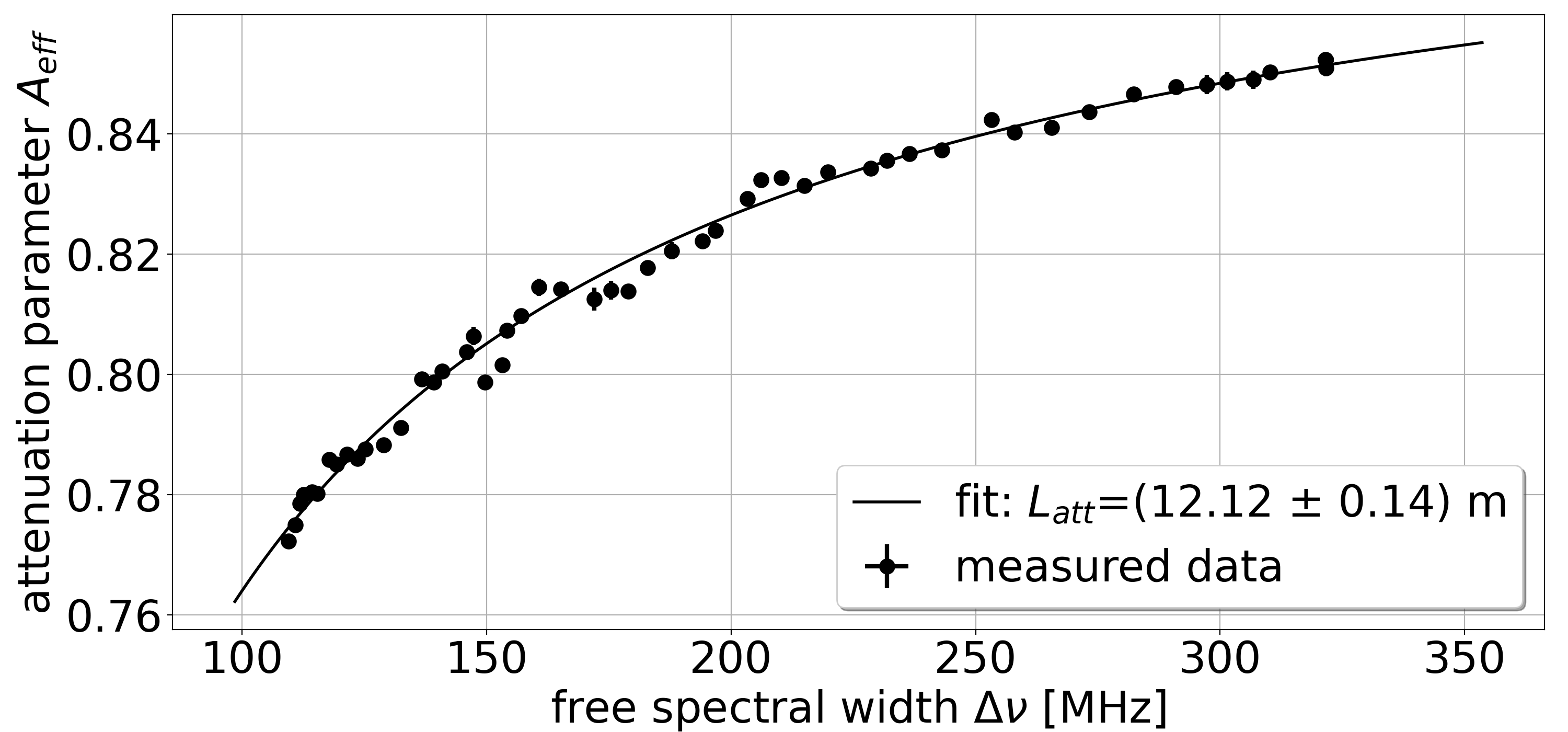}
	\caption{Measurement of the attenuation length $\Latt$ of \ac{LAB} manufactured by Sasol at the wavelength $\lambda=\unit[420]{nm}$. The attenuation length was estimated to be $\Latt=\unit[(12.09\pm0.14)]{m}$.}
	\label{fig:LAB_VLC}
\end{figure}

The group velocity $v_g$ and the attenuation length $\Latt$ of \ac{LAB} manufactured by the company Sasol were measured at two wavelengths $\unit[420]{nm}$ and $\unit[430]{nm}$ and the results are shown in 
Tab. \ref{tab:VLC_LAB_results}. 
The corresponding high precision measurement of \ac{LAB} at $\lambda=\unit[420]{nm}$ is illustrated in  Fig. \ref{fig:LAB_VLC}, considering a total of 51 different resonator lengths. A reference measurement with air at the largest resonator length was used to determine the group velocity $v_g$ in the sample (cf.  \ref{sec:Latt_uncert}). 
The attenuation length $\Latt$ was derived according to~\ref{eq:A_eff} using the least squares fitting method.
The group velocity $v_g$ was determined with an average relative uncertainty of $\sim\unit[0.03]{\%}$ and the attenuation length $\Latt$ was determined with an uncertainty of less than $\unit[2]{\%}$ for 51 different lengths.

\begin{table}[hbt]
		\centering
		\begin{tabular}{|c||c|c|}
			\hline
			\rule{0pt}{15pt}wavelength $\lambda$ [nm] &$v_g/c$ [\%]&$\Latt$ [m]\\ 
			\hline\hline 
			\rule{0pt}{15pt}$420\pm1$ &$63.502\pm0.013$&$12.09\pm0.14$\\
			\hline
			\rule{0pt}{15pt}$430\pm1$  &$63.777\pm0.023$&$14.67\pm0.36$\\ 
			\hline
		\end{tabular}	
		\caption{\ac{CELLPALS} results of the group velocity $v_g$ and the attenuation length $\Latt$ of \ac{LAB} manufactured by the company Sasol at two wavelengths $\lambda=\unit[420]{nm}$ (cf. figure \ref{fig:LAB_VLC}) and $\lambda=\unit[430]{nm}$.}
		\label{tab:VLC_LAB_results}
\end{table}

\subsection{Measurements of LAB from different purification stages}
\label{subsec:LAB_measurements}

This section presents the results of three samples from the purification pilot plant for the \ac{JUNO} scintillator from 2023~\cite{PurificationPlantJUNO}. $S_1$ and $S_2$ are \ac{LAB} samples from different
purification stages, $S_3$ is a sample of the \ac{LS} after mixing $\unit[2.5]{g/l}$ of the fluor \ac{PPO} and $\unit[3]{mg/l}$ of the wavelength shifter \ac{Bis-MSB} to the purified \ac{LAB}~\cite{PurificationPlantJUNO}.

\begin{table}[hbt]
		\centering
		\begin{tabular}{|c||c|c|}
			\hline
			\rule{0pt}{15pt} sample&$v_g/c$ [\%]&$\Latt$ [m]\\ 
			\hline\hline 
			\rule{0pt}{15pt}$S_1$ (LAB) &$63.727\pm0.024$&$21.05\pm0.41$\\
			\hline
            \rule{0pt}{15pt}$S_2$ (LAB) &$63.579\pm0.027$&$23.58\pm0.68$\\
			\hline
            \rule{0pt}{15pt}$S_3$ (purified LS) &$63.663\pm0.021$&$18.23\pm0.41$\\
			\hline
		\end{tabular}	
        \caption{Results of the group velocity $v_g$ and the attenuation length $\Latt$ for different purification stages of \ac{LAB} and the \ac{LS} derived with the \ac{CELLPALS} method at $\lambda=\unit[(425\pm1)]{nm}$.}
		\label{tab:LAB_result}
\end{table}
    
\begin{figure}[htb]
	\centering
	\includegraphics[width=\picsize]{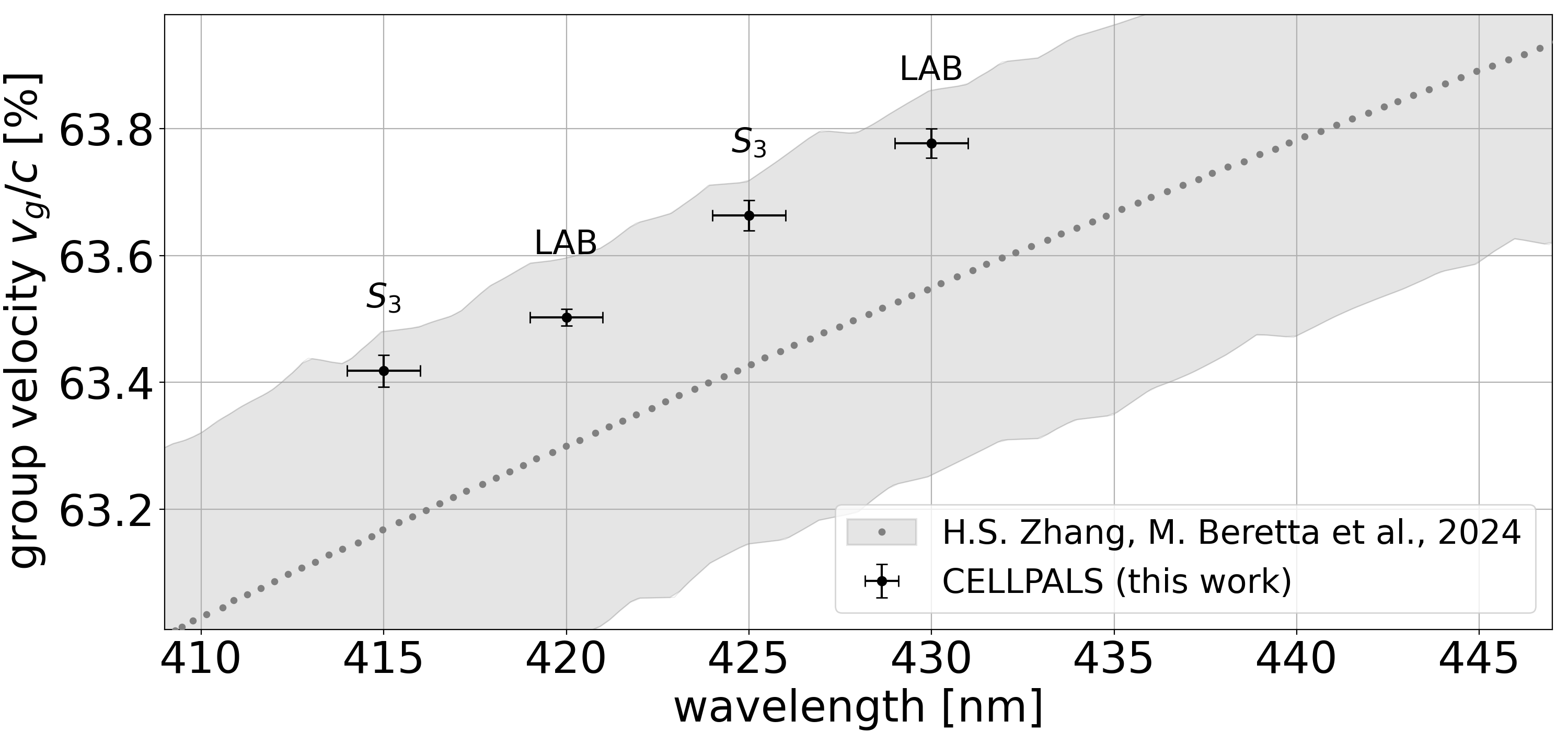}
	\caption{Comparison of the group velocity of sample $S_3$ (cf. table \ref{tab:LAB_result}) and \ac{LAB} from Sasol (cf. table \ref{tab:VLC_LAB_results}) with the group velocity in \cite{RefractiveIndexBeretta}, which was calculated from a wavelength-dependent measurement of the refractive index in the \ac{JUNO} \ac{LS}.}
	\label{fig:v_g_Beretta}
\end{figure}

In  Fig. \ref{fig:v_g_Beretta} the results of the group velocity of sample $S_3$ at the wavelengths $\unit[415]{nm}$ and $\unit[425]{nm}$ (cf. 
Tab. \ref{tab:LAB_result}) and of \ac{LAB} from Sasol at $\unit[420]{nm}$ and $\unit[430]{nm}$ (cf. 
Tab. \ref{tab:VLC_LAB_results}) were compared with the results in~\cite{RefractiveIndexBeretta}, where the group velocity $v_g$ is calculated from a wavelength-dependent measurement of the refractive index in the
\ac{JUNO} \ac{LS}. The results with \ac{CELLPALS} are in agreement with the calculated error band from~\cite{RefractiveIndexBeretta}, but are much more precise. 

\section{Conclusion and Outlook}
\label{sec:Conclusion}

\ac{CELLPALS} was developed as a method for precise attenuation length measurements for highly transparent media, that does not require a complex setup with a long sample holder. In direct comparison to the UV--Vis method, the \ac{CELLPALS} measurements with three fixed cavity lengths provided significantly more precise results with an average relative uncertainty of $\sim\unit[7]{\%}$ (cf. 
Tab. \ref{tab:FiveSamples_Latt}). The achievable precision of the attenuation length with the \ac{CELLPALS} method can be reduced to $\sim\unit[2]{\%}$ if more resonator lengths are used with a variable-length cavity (cf.  \ref{sec:Latt_uncert}). In addition, the UV--Vis method reveals previously unknown effects that lead to a systematic underestimation of the attenuation length for very transparent media (cf.  \ref{subsec:performance_measurement}). This is also confirmed by the comparison with the results obtained with the \ac{CELLPALS} method in 
Tab. \ref{tab:FiveSamples_Latt}. 

In terms of precision, the performance of \ac{CELLPALS} is comparable to the methods using a vertical tube~\cite{PALM,Latt_method1,Latt_method2,Latt_method3,Latt_method4} which is around $\sim\unit[2.5]{\%}$. However, due to its
compact design, \ac{CELLPALS} is more flexible in terms of exchanging samples and its installation location and requires less than $\unit[1]{litre}$ of sample for one measurement. In addition, the \ac{CELLPALS} method enables
the determination of the group velocity with a relative uncertainty of $\sim\unit[0.03]{\%}$ (cf.  \ref{sec:vel_uncert}), which is not provided by any other method for determining the attenuation length.

\ac{CELLPALS} supports the measurement of the attenuation length and the group velocity of any transparent liquid (like \ac{LS} or water) for a user-defined wavelength of the laser diode. The only requirement is an amplitude-modulated intensity up to $\sim\unit[350]{MHz}$ to determine the free spectral width of the resonator. For strong deviations from $\lambda=\unit[430]{nm}$, individual optical components such as the resonator mirrors or the beam splitter may have to be replaced. 

The \ac{CELLPALS} method still poses some challenges, as the data for some resonator lengths show deviations that are larger than the corresponding fit errors (see  Fig. \ref{fig:LAB_VLC}), which are probably related to the collimation of the measured intensity behind the cavity. As discussed in  \ref{sec:stat_fit_params} incomplete collimation leads to characteristic deviations in the amplitude ratio $\amp(\nu)$ and the phase shift $\phase(\nu)$ (cf.  Fig. \ref{fig:Sim_A}) and can thus be identified very clearly. To avoid such deviations, a larger sensitive area of the \ac{APD}s used would be useful in the future.

\appendix
\section{Intensity inside the cavity}  
\label{sec:I_out}

By substituting $J=IT^2\exp(-{L}/{\Latt})$ and $\tilde{t}(t)=t-\tau/2$ and by defining the attenuation parameter
\numberwithin{equation}{section}
\setcounter{equation}{0}
\begin{equation}
 \Aeff\equiv e^a\quad\text{with}\quad a=\ln(R^2)-\frac{2L}{\Latt}\,,\label{eq:Aeff}
\end{equation}
the total intensity in~\ref{eq:I_sum} (cf.  \ref{sec:Mechanism of CELLPALS technique}) is reduced to
\begin{equation}
 I_{\text{out}}(t)=Je^{i\omega\tilde{t}}\cdot\sum_{n=0}^{\infty}\left(e^{a-i\omega \tau}\right)^n\,.\label{eq:I_out_1}
\end{equation}
With $z=a-i\omega\tau~\in~\mathbb{C}$ the sum corresponds to a geometric series with the solution
\begin{equation*}
 Z=\sum_{n=0}^{\infty}\left(e^z\right)^n=\frac{1}{1-e^z}~\in~\mathbb{C}\quad\text{if}\quad \left\lvert  e^{z}\right\rvert  <1\,,
\end{equation*}
which is true due to $\Aeff<1$ (cf. Eq. \ref{eq:Aeff}). Therefore, Eq. \ref{eq:I_out_1} can be calculated to
\begin{equation}
 I_{\text{out}}(t)=Je^{i\omega\tilde{t}}~ Z\quad\text{with}\quad Z=\frac{1}{1-\Aeff\cdot e^{-i\omega\tau}}~\in~\mathbb{C}\,.\label{eq:I_out_2}
\end{equation}
By using the polar form $Z=|Z|\exp(i\phi)~\in~\mathbb{C}$ and by substituting $J$ and $\tilde{t}(t)$ back, the total intensity $I_{\text{out}}(t)$ in~\ref{eq:I_out_2} can be reduced to a simple oscillation
\begin{equation*}
 I_{\text{out}}(t)\equiv\tilde{I}(\nu)\exp(i\left[2\pi\nu t -\Delta\phi(\nu)\right])
\end{equation*}
with the real frequency-dependent amplitude
\begin{equation}
 \tilde{I}(\nu)=\frac{IT^2\exp\left(-\frac{L}{\Latt}\right)}{\sqrt{\Aeff^2-2\Aeff\cos(2\pi\nu\tau)+1}}~\in~\mathbb{R}\label{eq:App_Amp}
\end{equation}
and the real phase shift
\begin{equation}
 \Delta\phi(\omega)=\arctan\left(\frac{\Aeff\sin(2\pi\nu\tau)}{1-\Aeff\cos(2\pi\nu\tau)}\right)+\pi\nu\tau~\in~\mathbb{R}\,.\label{eq:App_Phase}
\end{equation}

\section*{Acronyms}
\begin{acronym}[CELLPALS]
	\acro{CELLPALS}{Cavity Enhanced Long Light Path Attenuation Length Screening}
    \acro{JUNO}{Jiangmen Underground Neutrino Observatory}
    \acro{LAB}{linear alkylbenzene}
	\acro{PPO}{2,5-Diphenyloxazole}
    \acro{Bis-MSB}{1,4-Bis (2-methylstyryl) benzene}
	\acro{LDs}{laser diodes}
    \acro{LS}{liquid scintillator}
	\acro{APD}{avalanche photodetector}
	\acro{PTFE}{polytetrafluoroethylene}
\end{acronym}

\section*{Acknowledgments}
We are very grateful for the support with the UV-Vis measurements at the TU Munich. In particular, L. Oberauer, H. Steiger and M. R. Stock have provided us with their help and have also been involved in many helpful discussions.
This work is supported by the DFG Research Unit FOR 5519 "Precision Neutrino Physics with JUNO".

\bibliographystyle{elsarticle-num}
\bibliography{mybibfile}

\end{document}